\documentclass[10pt, reqno]{amsart}
\usepackage{amsmath, amsthm, amscd, amsfonts, amssymb, graphicx, color}
\usepackage[bookmarksnumbered, colorlinks, plainpages]{hyperref}

\makeatletter \oddsidemargin.9375in \evensidemargin \oddsidemargin
\theoremstyle{definition}

\theoremstyle{remark}

\numberwithin{equation}{section}

\begin{document}

\title[A Historical View On The Maximum Entropy]
{\emph{A Historical View On The Maximum Entropy}}


\author[Fallah et al.]{Seyedeh Azadeh Fallah Mortezanejad}



\maketitle
\begin{center}
{\it School of Automotive and Traffic Engineering, Jiangsu University, Jiangsu, China.}
\end{center}

\begin{abstract} How to find unknown distributions is questioned in many pieces of research. There are several ways to figure them out, but the main question is which acts more reasonably than others. In this paper, we focus on the maximum entropy principle as a suitable method of discovering the unknown distribution, which recommends some prior information based on the available data set. We explain its features by reviewing some papers. Furthermore, we recommend some articles to study around the generalized maximum entropy issue, which is more suitable when autocorrelation data exists. Then, we list the beneficial features of the maximum entropy as a result. Finally, some disadvantages of entropy are expressed to have a complete look at the maximum entropy principle, and we list its drawbacks as the final step.
\end{abstract}

\section{\bf Introduction }
The entropic principle is used in a variety of fields of study, such as Statistics, Mathematics, Physics, Economics, etc. Shannon $ (1948) $ expressed first the concept of entropy. After that, Jaynes $ (1957a) $ declared the sense of maximum entropy. In the following years, it has been perused by some statisticians like Kagan et al. $ (1973) $, Shore and Johnson $ (1980) $, and Kapur $ (1989) $. The entropy liaises straightly with uncertainty. In other words, if entropy increases, the uncertainty measure will also do. The maximum likelihood concept is used when some model parameters are unknown, and the distribution is determined. Therefore, it cannot be applied to all estimation problems. So, if the distribution is obscure, the maximum likelihood method is incapable. Thus in this situation, maximum entropy is an effective way that does not need any assumptions on the prior distribution. Moreover, it works outstanding when the sample size is small. This property is very efficient because sometimes, the process cannot be sampled, for instance, sampling is financial consumption and ruin, or it takes lots of time. Hence, maximum entropy appears as a contributory method for researchers. \\
General maximum entropy is used when the number of unknown parameters is large. In several regression models, there is a basic assumption of error normality distribution. If errors are not fitted properly in a model, the researchers are in trouble. Thus, the general maximum entropy is practical in these conditions because the distribution of errors is not required, and it can be unknown. Another advantage is that outlying data cannot affect the estimation of general maximum entropy.\\
So many articles have presented the application of maximum entropy. For example, Autchariyapanitkul et al. $ (2017) $ studied the rice farmer's information. They found the maximum entropy density as a substitute for maximum likelihood. Tibprasorn et al. $ (2017) $ approximated the unknown parameters in panel data of macroeconomics by general maximum entropy because they had a limitation of data, which was an obstacle in their research. Some scholars like Parveen and Arora $ (2017) $, Ayusuk and Autchariyapanitkul $ (2017) $, and Kreinovich and Sriboonchitta $ (2017) $ got the help of this principle to predict the feature of different processes, which is elucidated more in one of the following sections. Fallah Mortezanejad et al. $ (2019) $ worked on capability indices in statistical quality control. They used the maximum entropy concept to find out the unknown distribution of manufacturing processes. Since in classic methods, there is a strong assumption of normality, they cannot detect small shifts in processes. Those distributions are not Normal, so these methods do not work accurately. Thus they applied the help of the maximum entropy principle to deal with this problem. Their constraints were according to a suitable capability index. Mortezanejad et al. $(2019)$ mixed two concepts of the maximum entropy and copula function to cope with the bivariate data set. Since the maximum entropy concept is a suitable way to apply prior information, and the copula function is a good way to save the original dependency between the data set, hence the introduced density is worth working with.\\
The organization for this article is: In section \ref{Sc2}, we present the basic concept of maximum entropy. In section \ref{Sc3}, we explain the advantages of the maximum entropy principle as well. In section \ref{Sc4}, we have a brief look at the papers whose scopes are according to the maximum entropy. In section \ref{Sc5}, we briefly state the weaknesses of maximum entropy in the form of a list and refer to some articles too. In section \ref{Sc6}, We offer a summary of the conclusion of the maximum entropy properties.

\section{\bf The maximum entropy concept}\label{Sc2}
The maximum entropy concept determines the most general distribution by some intended constraints. In this regard, the constraints influence choosing a distribution via this principle. These constraints can contain moment conditions or necessary qualifications based on the researcher's needs. As we mentioned before, Shannon $ (1948) $ expressed entropy $ H(f) $ of a continuous random variable $ X $ with density function $ f_X(.) $ on support set $ \mathbb{D} $ as below:
\begin{equation*}
  H(f) = - \int_{\mathbb{D}} \log f(x) dF(x),
\end{equation*}
which  is famous as Shannon entropy and was first introduced by Jaynes $ (1957a,1963) $ based on Shannon entropy. $ H(f) $ should be maximized concerning $ f $. Moreover, the distribution function $ f $ should satisfy the mentioned constraints. Our problem which has to be solved is:
\begin{displaymath}
\left\{\begin{array}{ll}
\max H(f), \\
E(\zeta_i|G) = \alpha_i,~~i=1, \ldots,k , \\
E(\zeta'_i|G) \leq \alpha'_j,~~j=1, \ldots,k'. \\
\end{array}\right.
\end{displaymath}
$ G $ is the true distribution function of process data estimated by the experimental selection of $ F $ via maximum entropy. $ \zeta_i $ for $ i=1,\ldots, k $ are some functions of moments. Some essential information is subjoined to the equation by $ \zeta_j ' $ for $ j=1, \ldots, k' $. The amounts of $ \alpha_i ,~ i=1,\ldots,k $ and $ \alpha_j', j=1,\ldots,k' $ are known and gained from the experimental information calculated from the desired process. For instance, assume $ EX= \mu $ and $ EX^2=\mu' $, which $ \mu $ and $ \mu' $ are known constraints on $ \mathbb{R} $. The maximization problem is:
\begin{displaymath}
\left\{\begin{array}{ll}
\max \left( - \int _{\mathbb{R}} \log f(x) dF(x)\right), \\
\int_{\mathbb{R}} dF(x)=1 , \\
\int_{\mathbb{R}} xdF(x)=\mu, \\
\int_{\mathbb{R}} x^2 dF(x)=\mu'. \\
\end{array}\right.
\end{displaymath}
The answer to the above equation is a Normal distribution with mean $ \mu $ and variance $ \mu'-\mu ^2 $. In the next section, we discuss supreme applications of the entropy principle, and then we explain some articles which have used maximum entropy in practice.

\section{\bf The application of entropy principle}\label{Sc3}
Here we list some advantages of maximum entropy and general maximum entropy. The first list is about maximum entropy, and it contains a brief explanation of some related papers as below:
\begin{enumerate}
  \item A maximum entropy distribution is an experimental choice of the true and most unbiased distribution. Penfield $ (2003) $ has presented the maximum entropy principle as the most unbiased distribution.
  \item It is a strong way to deal with multicollinearity and some ill-prosed problems such as data limitations and incomplete data. Yamaka et al. $ (2017) $ applied general maximum entropy expressed by Golan et al. $ (2017) $ for estimating the quantile regression model of capital assets pricing because of its power in multicollinearity situations.
 \item Prior distribution can be gotten parametric in quantile regression by using maximum entropy. The least-squares method is used for approximating unknown parameters in regression models with basic Normal assumptions. Yamaka et al. $ (2017) $ handled a quantile regression by maximum entropy as a nonparametric estimation of probability distribution functions and measurement uncertainty. Furthermore, when the dependence degree between variables and their effects increases on each other, the conditional entropy decreases until the lower bound. The increment or decrement of this information varies based on adding or removing variables. This content can be compared with a forward selection of variables in regression analysis according to the partial correlation in the multivariate model. Singh $ (2013) $ subjoined analytical derivation on multivariate probability distributions with the entropy concept, which is rigorous even with the marginal distributions of variables. This problem can be dealt with copula functions, which were introduced first by Sklar $ (1959) $. Mortezanejad et al. $ (2019) $ mixed the maximum entropy principle and copula function to save the dependency between variables on the estimated distribution.
  \item It also is reasonable for a small sample size. Leurcharusmee et al. $ (2017) $ studied child-gender preference with respect to the previous child and mothers' education in Thailand. General maximum entropy was a strong method for priority checking between their variables. The entropy principle has been recognized as a suitable way of dealing with multivariate cases. They have compared classical logit and entropy to estimate unknown parameters, which resulted in the betterment of the entropy principle because of the lower amount of standard errors and needleless of large sample size.
  \item It has lower standard errors compared with the logit method in a miniature sample size. Leurcharusmee et al. $ (2017) $ analogised the entropy method and logit. The power of the entropy was its lower value of standard errors.
  \item It provides a prediction of the future and has the least surprising in prediction terms. Conrad $ (2004) $ declared that the distribution which satisfies desired constraints has a lower surprising in predictions.
 \item It contains available information and is uniformly the most consistent distribution concerning prior information. Yang $ (1997) $ said that the maximum entropy has lower prejudice than all other consistent distributions respect to the intended constraints. This principle does not specify any domain for random variables. He has merged the entropy concept with Pearson's system to determine the range of random variables. The maximum entropy performs better if any assumptions of their domains are not available. Another mentioned advantage was applying some desired constraints on it. Ebrahimi et al. $ (2008) $ declared that the multivariate distribution can be modelled based on maximum entropy using moment constraints, and it does not have the problem of finding distributions. They have added consistently available information on the result distribution as an advantage of the maximum entropy principle, which is true for multivariate cases, Soofi et al. $ (1995) $. Rahman and Majumdar $ (2003) $ found the multivariate density of a finite set via maximum entropy. The data matrix contained some prior information.
  \item Any assumptions on prior joint distribution are not necessary. It is suitable for quantitative and qualitative variables. In addition to the moment conditions, any other required constraints can be applied to its result distribution. The aim of Ayusuk and Autchariyapanitkul's $ (2017) $ paper was prediction influencing factors in returning a tourist to an area in Thailand. The logistic regression model was applied for this goal. The general maximum entropy was exploited for the estimation of the unknown parameters. This method is much more tenuous than the maximum likelihood when the sample size is small. A consistent distribution has to be attained by the maximum entropy principle, which assures the previous information and required constraints. It makes lower surprising than other distributions. As the advantages of maximum entropy distribution, it can be noted that it consists of all available information. There is no need for any assumption about the joint distribution of the data. So, it is suitable for quantitative and qualitative data, and in addition to moments, any other constraints can be added to the desired distribution. The reason for maximum entropy selection is similar to the maximum likelihood, which solves the problem by maximizing the likelihood function. The maximum entropy exerts the Lagrange function, and required constraints can be added to the result. General maximum entropy works better than the likelihood method because of its lower amount of $ MSE $. They have used different sample sizes $ 200 $ and $ 400 $. Both of them acted well when the sample size was $ 400 $, but when it was decreased to $ 200 $, the entropy principle worked superior, Ayusuk and Autchariyapanitkul $ (2017) $.
  \item It does not require parametric assumptions. Campbell $ (1999) $ pointed to Soofi $ (1992) $ and Golan et al. $ (1996) $, and expressed that any parametric assumption is not required in maximum entropy. Zhang $ (2009) $ has compared the maximum entropy and maximum likelihood method. Their outcomes were not equal. The privilege of maximum entropy was no need for parametric assumptions.
  \item This distribution consists of small numbers of unknown parameters, which have to be estimated. Park $ (2007) $ used the maximum entropy concept to estimate portfolio weights in asset allocation problems because it decreased the number of estimated parameters.
  \item It is a general case of machine learning. Jaynes $ (1957a) $ and shin $ (2009) $ have gotten the conditional probability distribution with maximum entropy in machine learning.
  \item It is a logical way to classify heterogeneous information. Patnaparkhi $ (1998) $, Manning and Schutze $ (1999) $, and shin $ (2009) $ applied maximum entropy to manage various and heterogeneous information in their study.
  \item It has the most amount of uncertainty. Maximum entropy was introduced as a consistent distribution by Penfield $ (2003) $ and is well-behaved for uncertainty on the multivariate cases in ill-posed conditions, and has a unique solution for the problem. Rahman and Majumdar $ (2004) $ represented the usage of maximum likelihood in some cases that are too austere. Thus, it can be reasonable to apply the maximum entropy concept to the multivariate quandaries.
  \item It is a robust tool in image processing. Skilling and Gull $ (1984a) $ and Skilling used maximum entropy in image restoration with a different type of data. Its application is usability in radio astronomical interferometry because it decreases certainty.
  \item It is proper for noisy data. Gzyl $ (1998) $ and Golan and Dose $ (2001) $ are confronted with noisy data via maximum entropy based on Shannon $ (1948) $.
  \item The goodness-of-fit test can be done for multivariate data using maximum entropy, which had been studied in Rahman and Majumdar $ (2004) $.
\end{enumerate}
The general maximum entropy is more universal than maximum entropy and has some similar properties. The below list is about the advantages of general maximum entropy:
\begin{enumerate}
  \item The general maximum entropy is suitable when the number of unknown parameters is high. Khiewngamdee et al. $ (2017) $ have easily estimated several unknown parameters using general maximum entropy.
  \item It exerts all information that data gives. Khiewngamdee et al. $ (2017) $ and Chinnakum and Boonyasana $ (2017) $ have utilized all previous information in their study.
  \item Outlying data cannot affect their results. This application of general maximum entropy has been discussed in Khiewngamdee et al. $ (2017) $.
  \item It is an appropriate way to estimate parameters without any extra assumption on errors. Also, unknown probabilities and errors can be jointly estimated. General maximum entropy allows us to approximate unknown probabilities and errors jointly, and any extra assumption is not needed. Therefore, any other consistent distributions have lower entropy or uncertainty. In classical technics, strong assumptions are required to have unique results. The entropy principle solves this problem, and any prior information can be added to the problem whose result is uniformly the most consistent distribution, Jaynes $ (1957a)$, Campbell $ (1999) $. Golan et al. $ (1996) $ expressed that any presumptions about errors were not necessary. Golan et al. $ (2000) $ used this application and noted that it is a robust and efficient method compared with maximum likelihood.
  \item Bootstrap is not needed via general maximum entropy. Yamaka et al. $ (2017) $ and Navapan et al. $ (2017) $ said the general maximum entropy as a method that is not used Bootstrap.
  \item It is rational for linear and nonlinear regression when the sample size is small, Navapan et al. $ (2017) $. Without any doubt, bank performance assessment concerns people like bank directors to schedule for the future. Navapan et al. $ (2017) $ exerted the classical stochastic frontier model and efficiency stochastic frontier model. The advantages of general maximum entropy were motioned that it does not need the bootstrap method, the high sample size in linear and nonlinear regression models, and any extra classical assumption about error distribution. So, it is fitted to models with non-Normal errors.
  \item Any parametric assumptions about the error distribution are not required. Golan et al. $ (1998) $ noted that any extra assumptions about the basic distribution was not needed, unlike maximum likelihood. The aim of Golan et al. $ (1999) $ was to estimate a set of unknown parameters by imposing all available information without any prior assumption on the basic distribution. So, they applied general maximum entropy. Their coefficients had lower errors than other estimators. Glennon and Golan $ (2003) $ expressed that it is semiparametric and is useful for limited data. Also, it allows us to utilize prior information, so it is more suitable than the maximum likelihood. Moreover, they emphasized that any assumptions about the distribution are not desired in the entropy principle.
  \item It is fitted to models with non-Normal errors. General maximum entropy does not require any assumptions on error distribution, but in some other methods, it is assumed to be Normal, which is not always true. Thus the entropy principle is fitted when errors have a non-Normal distribution, which was used by Khiewngamdee et al. $ (2017) $ and Yamaka $ (2017) $.
  \item Unknown parameters estimated by general maximum entropy have lower variance than logit, probit, and ME-logit manners. Campbell $ (1999) $ declared that the entropy concept was a new and appropriate method in economics. Also, he compared estimators of maximum entropy and general maximum entropy with estimators of linear regression, binary choice, multinomial regression models, censored data, and truncated regression models. The result was the superiority of the entropy principle. Denzau et al. $ (1989) $, Soofi $ (1992) $, and Golan et al. $ (1997) $ worked on this concept of entropy. Another consequence of Campbell $ (1999) $ was that: if the width of errors is increased, variances of general maximum entropy estimators are decreased, and biases are increased. Golan et al. $ (2000) $ said that maximum entropy is a particular mode of general maximum entropy, which does not have any weight on noise components and is similar to maximising the logistic likelihood function. The general maximum entropy was more flexible and efficient than the maximum likelihood logit method. Golan et al. $ (2001) $ remarked that it is more consistent than maximum likelihood and least square, so it has lower variance.
  \item It uses non-sample information. Campbell $ (1999) $ said that non-sample information could be imposed on general maximum entropy estimators in linear regression.
  \item It imposes support points to the unknown parameter estimations. Support points are involved in general maximum entropy results, but it is difficult with other methods.
  \item It is useful for a tiny and unbalanced sample size. Golan et al. $ (1999) $ discussed that general maximum entropy prepares more stable estimators in a small sample size. It has been introduced for small and ill-posed samples by Golan et al. $ (2001) $. Kullback $ (1959) $, Levine $ (1980) $, Shore and Johnson $ (1980) $, Skilling $ (1989) $, Csiszar $ (1991) $, Golan et al. $ (1998) $, and Wu et al. $(2006) $ reviewed maximum entropy and its applications and told that it is a special case of general maximum entropy without any noise weight components and other observations.
  \item The estimator bias is lower than ME-logit. Maximum entropy has been expressed as an approximate method compared with ordinary least squares when the regression range increases whose bias is always lower than other estimators, Leung and Yu $ (1996) $, Golan et al. $ (2001) $.
  \item It works well with Mont Carlo experiments, Yamaka et al. $ (2017) $
  \item It is consistent concerning prior information. General maximum entropy was presented as a consistent distribution estimation to the convex assumption by Golan and Perloff $ (2002) $, Golan et al. Golan et al. $ (1996) $, and Golan et al. $ (1998,~ 2001) $.
  \item It has a lower amount of $ MSE $. Golan et al. $ (1999) $ showed that general maximum entropy has lower $ MSE $ than ordinary least squares, Heckman's method, maximum likelihood, and Powell's ways. Golan et al. $ (1997) $ and Golan et al. $ (2000) $ said that it has lower empirical $ MSE $ than maximum likelihood. Campbell $ (1999) $ and Golan et al. $ (2001) $ declared that $ MSE $ of general maximum entropy estimators are always lower than any other estimators and are more consistent than maximum likelihood.
  \item Its power in prediction is more than the Maximum likelihood. Golan et al. $ (2000,~2001) $ worked on prediction, and their result was the superiority of general maximum entropy to the probit method.
  \item The simplicity of calculation is another benefit. Golan et al. $ (1998) $ utilized the general maximum entropy simplicity of calculation to approximate several unknown parameters, and imposed different inequalities on constraints.
\end{enumerate}

\section{\bf Perusing some papers on the benefit of maximum entropy}\label{Sc4}
In the previous section, we focus on the properties of the maximum entropy and general maximum entropy. Here, some interesting papers are presented, which dissolved variant knots by the maximum entropy principle.
\subsection*{Maximum entropy as a feasible way to describe joint distribution in expert systems}
Let a system define with different properties by an expert. The probability functions of these reports have to be known. Furthermore, their joint distribution is needed to be determined. The problem is that some knowledge is available, and their distributions are unknown. Dumrongpokaphan et al. $ (2017) $ solved this obstacle via maximum entropy distribution. The corresponding entropy is maximized by various unknown parameters. They found a probability distribution function that satisfied desirable conditions that were taken by the expert.
\subsection*{Entropy as a measure of average loss of privacy}
The lack of someone's information for others means her privacy. Some extra questions should be asked to know her personal information. Longpre et al. $ (2017) $ measured the degree of privacy by the average number of Yes or No questions. Shannon entropy was fitted in such a case. They have illustrated that entropy is not a good measure of absolute privacy loss, but it is suited for the mean of privacy lack, which aimed to decline uncertainty. Privacy can be measured by entropy distribution, which became famous as the average number of Yes or No questions that should be queried to give enough knowledge. Thus, the entropy principle can be used in uncertainty concepts.
\subsection*{Probabilistic graphical models follow directly from maximum entropy}
Probabilistic graphical models are an efficient manner in machine learning. The maximum entropy method is significant in these models. The idea of Ly et al. $ (2017) $ was maximizing the absolute entropy to achieve the most normalization. Partial information was accessible, and several probability functions satisfied their knowledge. Therefore, the most reasonable distribution has the most uncertainty and entropy. Thus, among all probability distributions assuring favourable constraints, the distribution that has the most entropy should be selected.
\subsection*{How to get beyond uniform when applying MaxEnt to interval uncertainty}
Sriboonchitta and Kreinovich $ (2017) $ declared that the maximum entropy principle is more logical and fitter in practices than hit off in considered conditions. They have exerted a fuzzy technic to deal with uncertainty intervals.
\subsection*{Capital asset pricing model through quantile regression: An entropy approach}
There is the main difference between probability and entropy. A probability is a quantitative measure of an occurrence chance, but entropy is a measure of uncertainty for a random variable. Therefore, Yamaka et al. $ (2017) $ constructed a model for financial risk via the entropic principle. So, they maximized entropy information as a continuous function under three constraints:
\begin{equation*}
  f_{ME} = \arg \max - \int_{\mathbb{D}} f(y)\ln f(y) dy;
\end{equation*}
\begin{displaymath}
\left\{\begin{array}{ll}
\int f(y) dy =1, \\
E|Y-x\beta^{\tau}|=C_1 , \\
E(Y-x\beta^{\tau})=C_2, \\
\end{array}\right.
\end{displaymath}
where $ C_1 $ and $ C_2 $ are known constants. ADL distribution is assumed for the unknown distribution, but it is not always true. Hence, maximum entropy can solve the presented problem. Finally, general maximum entropy estimation showed better performance than classical least squares, maximum likelihood, and Bayesian estimations via Monte Carlo simulation.
\subsection*{An entropic structure in capability indices}
In classical quality control, the distribution of manufacturing processes is supposed to be Normal, but this assumption is not always correct. Fallah Mortezanejad et al. $ (2019) $ used the maximum entropy principle to find out the distributions of manufacturing processes as well. They applied a capability index to the constraints of maximum entropy to make sure that the processes with this distribution are in statistical control. Then, they made an updated form of the distribution based on its unknown parameters using the Kullback-Leibler information measure. Thus, by the aim of this article, there is no need for any pre-assumption on the distribution of processes.
\subsection*{Estimating release time and predicting bugs with Shannon entropy measure and their impact on software quality}
The concept of entropy is an information theory that measures first uncertainty. Parveen and Arora $ (2017) $ obtained a proper prediction model based on previous information and believed that prediction based on entropy was more accurate than any other estimators. Their method was to fit some suitable regression models to data via the entropic principle.
\subsection*{MaxEnt-based explanation of why financial analysts systematically under-predict companies performance}
Studies show that financial analysts disregard the prediction of business performance. Kreinovich and Sriboonchitta $ (2017) $ explained the maximum entropy solution for the existing problem. Among all favourite distributions, they selected the one which has the most entropy. The notable point was that all favourite distributions satisfied the significant constraints, but the one, which had the most entropy, was their desire.
\subsection*{Maximum entropy approach to interbank lending: Towards a more accurate algorithm}
Nguyen et al. $ (2017) $ studied minimizing the possible risk of the prediction of lending between several banks. A solution is to select the interbank rates which have the most entropy. In this situation, the marginal distribution and the maximum entropy of joint distribution were gained. Then, they selected the observations which have the most entropy:
\begin{equation*}
  \sum_{i,j} x_{ij} \ln (\frac{x_{ij}}{x_{ij} ^{(0)}}).
\end{equation*}
\subsection*{An empirical examination of maximum entropy in copula-based simultaneous equations model}
Choktaworn et al. $ (2017) $ worked on dependence data and made a model by entropic principle. Then, a copula function was applied for improving the model. This method did not need any prior assumption on the marginal distributions. Therefore, they got the margins via maximum entropy. So, they used the entropy concept instead of using parametric copula functions. This helped them to insert the available information into the result distribution. Hence, the entropy principle was exerted to the copula theory. They introduced the entropy-copula method as an appropriate method to model dependence data, especially for error terms. Maximum entropy distribution was uniquely determined by the Lagrange function and suitable constraints. Thus Spearman measure was used to measure the dependency among variables in an entropy-copula manner.
\subsection*{Portfolio optimization of entropy commodity futures returns with minimum information copula}
Tarkhamtham et al. $ (2017) $ checked out goods efficiency, which consumed energy by Garch and Egarch models. Also, they applied  copula functions to models that satisfied the constraints of the optimum energy consumption. Then a copula function was selected with minimum entropy. Moreover, vine copula was applied to measure the risk in a sample via minimum information copula. Thus, the dependency between variables was transferred to the copula function by putting moment constraints.
\subsection*{Joint dependence distribution of data set using optimizing Tsallis copula entropy}
In some research, the main problem is dealing with multivariate data sets. In these cases, there are two major questions, the first one is how to find the unknown distributions, and the second one is how to save the original dependency between different variables in a data set. Mortezanejad et al. $ (2019) $ worked on this issue using Tsallis and Shannon entropy. They combine two outstanding concepts of the maximum entropy and copula function to cope with the explained situation. A maximum copula entropy function was gotten in their paper. Then they used this function to get the unknown distribution of a bivariate data set. Their constraints consisted of Spearman's rho measure to transfer the dependency to the resulting density. They applied some surface plots and counterplots to show the results in some examples. In their article, there are different tables for the examples, which contained the Lagrange coefficients of maximum copula entropy functions. Furthermore, they compare Tsallis and Shannon entropy in the paper.
\subsection*{Coffee stochastic frontier model with maximum entropy}
A strong assumption on distribution is essential in classical maximum likelihood. So, unreliable interpretations result when the probability distribution is unknown. To deal with this obstacle, Khiewngamdee et al. $ (2017) $ used general maximum entropy for estimating the stochastic frontier model. Therefore, any additional probability distribution assumptions were not needed. Moreover, the general maximum entropy was useful when several parameters should be estimated. Also, it helped to reduce external variables in multiple cases. General maximum entropy allowed them to use prior information on parameters by putting the support bounds and some suitable consistent constraints. They declared the advantages of general maximum entropy that used the available information, cannot be affected by outlying data, and additional assumption on error terms is not required. They compared two methods maximum likelihood, and general maximum entropy. Finally, they offered the usage of the entropic principle.
\subsection*{The sample selection model: Application on the farmer's decision of rice acreage}
Yamaka et al. $ (2017) $ expressed that estimating parameters of a model is inconsistent when the distribution is unknown. To deal with this obstacle, primal general maximum entropy was applied with some constraints on sample selection. Also, they used multivariate response variables for the approximation of the model's parameters. In practice, error distribution is unknown, and residuals cannot be easily fitted to any distribution. General maximum entropy needs much lower assumptions on error distribution. So, it does not require any basic prescriptions for distribution. Thus, maximum likelihood is not suitable when it is passive. Then, its sample size was tiny, and the case study consisted of multivariate variables. So, the difficulty was dealing by the entropic principle.
\subsection*{Generalized information theoretical approach to panel regression kink model}
The objective of Tibprasorn et al. $ (2017) $ research was to approximate parameters in macroeconomic panel data via general maximum entropy. The usual problem is data limitations in this field of study. Further, the introduced method and an appropriate regression model solved the description of the relationship between explanatory and response variables. Their model was kink regression. They applied general maximum entropy to estimate parameters.
\subsection*{Maximum entropy quantile regression with unknown quantile}
A selection of quantile levels is the main problem of several researchers in the quantile regression model. Chokethaworn et al. $ (2017) $ worked on primal general maximum entropy for selecting unknown parameters such as quantile ones. The concept of entropy is a proper way to decrease different parametric assumptions in the survey.
\subsection*{Modelling Thailand tourism demand: A dual generalized maximum entropy estimator for panel data regression models}
Chinnakum and Boonyasa-na $ (2017) $ investigated factors that affected international tourist behaviour in Thailand. General maximum entropy was applied to find the parameters of the regression model. Its advantages were to work well in ill-prosed situations such as data limitations or incomplete data. They compared ordinary least-squared and general maximum entropy estimators. The entropic concept acted preferable because of its less $ MSE $. Furthermore, previous information and support of parameters have participated in an entropy manner to estimate parameters in the regression model.
\subsection*{Technical efficiency in rice production at farm level in northern Thailand: A stochastic frontier with maximum entropy approach}
Autchariyapanitkul et al. $ (2017) $ perused rice farmer's information and introduced maximum entropy as a substitute method for maximum likelihood. Its reason was a lower amount of $ MSE $ . Also, they compared standard errors in the frontier model with the conventional stochastic frontier model without the entropy.
\subsection*{Measurement and comparison of rice production efficiency in Thailand and India: An efficient frontier approach}
Sirikanchanarak et al. $ (2017) $ worked on general maximum entropy in the stochastic frontier model to investigate rice production in Thailand and India. Its root means square errors were lesser than in the traditional stochastic frontier model. Thus, the mentioned manner has been more fitted than the classical method.
\subsection*{Pension choices of senior citizens in Thailand: A multi-label classification with generalized maximum entropy}
Ruanto et al. $(2017) $ perused the information on international banks and used semiparametric general maximum entropy in their models. They concluded that this method was much more appropriate than other classical logit and probit methods, which are parametric. Logit and probit did not respond when the data distribution was not Normal or logistic. Furthermore, general maximum entropy has estimated parameters with fewer variances. Thus, it has strongly approximated them and preserved dependence among data.
\subsection*{Multivariate extensions of maximum entropy methods}
Justice $ (1985) $ studied solving problems of spectral maximum entropy estimation on a single channel to increase the science in this field of study. Clearly, all interests have been absorbed in the Burg algorithm because there were so many different residual matrices in recursive equations. There are lots of methods to deal with the obstacle, but they could not give a stable prediction. Finally, Strand $ (1977) $ and Morf et al. $ (1978) $ recommended a solution for single-channel. So, a multichannel extension of spectral maximum entropy was one of the favourable procedures without any heavy calculations and had a bit different manner. They answered some questions about multidimensional spectral maximum entropy. Eventually, the problem was decreased to optimum a convex function over a convex set with a finite dimension. Thus, maximum entropy worked as well as the Burg algorithm, but it did not annoy in the multivariate model. Also, Marcano et al. $ (2017) $ studied the algorithm and entropy principle.
\subsection*{On solutions to multivariate maximum $ \alpha $-entropy Problems}
The aim of Costa et al. $ (2003) $ was to introduce some properties of multivariate maximum entropy distributions in the general class whose name is Renyi's $ \alpha $-entropy on covariance constraints. First, they mentioned spectral invariance as an application of maximum entropy distribution used in some fields of study like pattern recognition, communication, independent components analysis, and inverse obstacles.

So many articles were perused on the beneficial application of maximum entropy as  Dickinson $ (1978) $, Dien and Thuy $ (2006) $, and Yu $ (2008) $. They are used in different fields of study like spline kernel and several papers on its convenient applications like Pillonetto and Nicolao $ (2011) $, Carli $ (2014) $, and Chen et al. $ (2016) $. In the following section, we focus on some drawbacks of the maximum entropy principle to complete our research.

\section{\bf Some weakness of maximum entropy}\label{Sc5}
In this paper, we focus on some attributes of the maximum entropy concept and study articles, which used these features. However, the maximum entropy principle has some disadvantages in practice. Fung et al. $ (2004) $ and Wallach $ (2004) $ pointed to its incompetence and talked about some dependence, which makes it hard to use. Hence, we prefer to discuss its weaknesses, which makes a complete review of maximum entropy, and some of them are listed below:
\begin{enumerate}
  \item A clear debility is the complexity and time-consuming for the Lagrange calculation of maximum entropy distribution when the constraints are a bit more. Therefore, Ingman and Merlis $ (1992) $ expressed this complication and added that the entropy method was unsuitable for signal processing. They applied that in neural networks and showed the entropy method is a proper way to deal with the invariant state in Hopfield net for estimating parameters, but the convolution annoyed them. Another hardness was the necessity of making an equivalent temperature. Also, Graf et al. $ (1988) $ and Marrian et al. $ (1989) $ studied the same field. Erdogmus et al. $ (2004) $ introduced a linear equation method to decrease the complicity of calculation for finding the Lagrange coefficients, but it had lower accuracy. Another suitable numerical way was inserted by Balestrino et al. $ (2006) $ for estimating the maximum entropy distribution, which could not approximate correctly too. Chen et al. $ (2010) $ used Newton's manner to solve a non-linear system of equations to find out the Lagrange, but this had two deficiencies. First, the calculation is too expensive because of having too many numerical integrals. Second, amounts of estimated Lagrange were too sensitive concerning the selection of primitive points. It is a good way when the number of moments constraints is limited. Furthermore, Ashcroft $ (2011) $ expressed the complicity of maximum entropy for finding peak width and said that this method is not precise enough. Tan and Taniar $ (2007) $ emphasized the estimation form to compute parameters of distributions.
  \item Another shortcoming of maximum entropy is the disability of measuring noise discussed in Lane $ (2008) $. Jaynes $ (1982) $ declared that the only appropriate solution for noisy data is the Bayesian approach. Pinna et al. $ (1993) $ studied some properties of maximum entropy and Acacia criteria in their investigation. They mentioned that the worth of the work would decrease if noisy data had been available because Acacia has a direct relationship with maximum entropy. Thus, it was improper for this type of data.
  \item Some papers like Djafari and Demoment $ (1988) $ introduced the maximum entropy in an uncreditable way, and it may repeat the curvature information several times. However, it was perused as a strong technic for retouching images with noisy and uncompleted data. It allows us to opt for a probability distribution based on prior information.
\item A big mistake of maximum entropy is too conservative and loses some interesting physical features. Kong and Lynn $ (1990) $ imported a method of measuring the X-ray spectroscopy according to maximum entropy, while positron beams had variable energy. That was weak because of using entropy principles to improve system resolution.
\item Hashimoto and Kobune $ (1985) $ used maximum entropy in directional spectral with three equivalent values for random motion waves. On the other hand, Hashimoto et al. $ (1994) $ declared its fault, and it is not a general way because it was limited to pre-arranged values.
  \item Another defect is to have the same entropy for two different images when their autocorrelation function is equivalent mentioned by Desoky and Hall $ (1990) $.
\item Maximum entropy has incompetence in using prior information discovered by Jedynak and Khudanpur $ (2005) $. It happened when there were not any constraints, prior information was not invariant or incorrectly giant for the interchange of variables. So, the numerical method cannot be helpful. Lane $ (2008) $ applied the MCMC algorithm in such situations.
\item Carbonell and Siekmann $ (2005) $ pointed out that maximum entropy does not answer when a system of constraints based on a non-closed set is available.
\end{enumerate}
\subsection*{A generalization of the entropy model for brand purchase behavior}
Kapur et al. $ (1984) $ had a study on economics, and one fault of maximum entropy was non-flexibility in estimating brand switching. The entropy approximation was based on market share and independent of product categories. It is reasonable that brand loyalty affected the probability of re-buying and was different from production categories. They expected various brand loyalty for some products such as cigarettes, cereal, gasoline, and soft drink.
\subsection*{Convolutions with correct sampling}
Magain et al. $ (1998) $ determined a disadvantage of maximum entropy in image processing because it used the total point spread function (PSF) in decomposing an image. The problem of sampling was the solution when each point stood on an exact pixel. They applied a finer pixel grid of modelling to increase the precision. The obligation of specifying positron by the user was another weakness of maximum entropy, which cannot be set with an algorithm. Another failure of this classical method named Richardson-Lucy was the dependency of the solution on zero points in an image.

\section{\bf Conclusion}\label{Sc6}
In the previous sections, we illustrate several articles that applied suitable properties of the maximum entropy principle to have superior results. They used the entropy concept in different fields of study like Economics, Physics, Communication theory, and Computer science, which shows the spread of its utilization. As mentioned, the general maximum entropy is for autocorrelated data. It has some convenient applications which can help researchers in their studies. Maximum entropy is a subset of general maximum entropy, and the major distinction is the loss of weight on noise components and any assumptions on errors. Briefly, we can say about the advantages of this method that:
\begin{itemize}
  \item Lower $ MSE $
  \item Any parametric assumptions and error distribution are not required
  \item Offered for ill-posed conditions
  \item Limited data
  \item Noisy data
  \item Unbalanced and small sample size, etc.
\end{itemize}
Thus, we provide some reasons for the maximum entropy usage, and some papers are presented as examples of its utilization.

The maximum entropy has some frailties in some fields of study. We peruse articles that mentioned its faults. The biggest problem is the complexity of the calculation with several numerical integrals. Moreover, this distribution is extremely influenced by prior information, which is used as constraints of the maximum entropy method. Hence, wrong information can change it and has a bad effect on results. This is not the deficiency of the entropy principle but is the defect of users.


\end{document}